# A Fuzzy Reward and Punishment Scheme for Vehicular Ad Hoc Networks

Rezvi Shahariar, Chris Phillips
School of Electronic Engineering and Computer Science, Queen Mary, University of London, London, England

*Abstract*—Trust management is an important security approach for the successful implementation of Vehicular Ad Hoc Networks (VANETs). Trust models evaluate messages to assign reward or punishment. This can be used to influence a driver's future behaviour. In the author's previous work, a sender-side based trust management framework is developed which avoids the receiver evaluation of messages. However, this does not guarantee that a trusted driver will not lie. These "untrue attacks" are resolved by the RSUs using collaboration to rule on a dispute, providing a fixed amount of reward and punishment. The lack of sophistication is addressed in this paper with a novel fuzzy RSU controller considering the severity of incident, driver past behaviour, and RSU confidence to determine the reward or punishment for the conflicted drivers. Although any driver can lie in any situation, it is expected that trustworthy drivers are more likely to remain so, and vice versa. This behaviour is captured in a Markov chain model for sender and reporter drivers where their lying characteristics depend on trust score and trust state. Each trust state defines the driver's likelihood of lying using different probability distribution. An extensive simulation is performed to evaluate the performance of the fuzzy assessment and examine the Markov chain driver behaviour model with changing the initial trust score of all or some drivers in Veins simulator. The fuzzy and the fixed RSU assessment schemes are compared, and the result shows that the fuzzy scheme can encourage drivers to improve their behaviour.

## I. INTRODUCTION

Vehicular Ad Hoc Networks (VANETs) can play a major role in the successful implementation of the Intelligent Transport System (ITS). However, the implementation of VANETs and ITS face many security threats concerning traffic events. There are many security approaches in state-of-the-art literature which aim to address these threats, though the completeness of these approaches is limited in thwarting both internal and external attacks. Attacks from authorized users can be curbed by a trust model [1-4]. However, each trust model has some limitations. Research to-date presents many models to evaluate the trust of vehicles and their messages. Trust approaches differ by their evaluation mechanism, and the infrastructures considered in the approaches. Some schemes evaluate only the trust of vehicles, whereas others evaluate the trustworthiness of messages. There are also some hybrid approaches which evaluate both the vehicles and messages. In this way, approaches isolate malicious vehicles from benign ones. Typically, the trustworthiness of relayed messages is evaluated using some measures and computational processes [5]. Once the malicious vehicles are identified, then it possible to limit or ignore their actions. To this end, some schemes also blacklist vehicles and/or drivers [6, 7]. Additionally, approaches incentivize trustworthy announcements (positive behaviour) to motivate vehicles to act honestly in the future [3, 6, 8]. Conversely, approaches punish mischievous behaviours to limit their future actions to avoid launching of future attacks [6, 9]. By arranging punishments to lower their trust score, drivers may feel guilty and be more careful about their future actions. In this way, the VANET can thwart attacks from authorized users by adopting a trust model which penalizes malicious activities and rewards benevolent behaviour. Even so, in some approaches [2, 3, 5, 10, 11], both reliable and unreliable vehicles can announce messages. The approach in [6] does not need any trust metric dissemination unlike these schemes [2, 4, 12] which require substantial trust data dissemination to verify an original announcement.

Trust evaluation can be performed at either the sender [6] and/or receiver side [2, 4, 10, 11, 13]. If receiver vehicles evaluate the trust of sender vehicles and/or messages, then the approach incurs additional delay and results in a higher communication overhead. Whereas, if a device in the sender vehicle can evaluate the trust, then there is typically no need to evaluate the trust of sender messages. Receivers do not need to rely on further communication with other sources (RSU, neighbour vehicles) for opinions or recommendation data.

A trust management framework is proposed in [6] which adopts sender-side trust evaluation inside a Tamper Proof Device (TPD) which is equipped onto every regular vehicle. This TPD is responsible for altering the trust of all drivers of a regular vehicle. In this approach, drivers receive rewards from announcements after the expiration of reward withhold timer. The accuracy of the message, responsiveness of the driver, and the distance travelled from the event location are used to calculate the reward or penalize a driver. The TPD updates the trust of a driver using a standard set of rules. The TPD does not know the truthfulness of a message unless it receives a report/complaint from a reporter vehicle about the announcement. The framework thus includes a collaboration procedure to determine the validity of a disputed event by an RSU. The dispute concerns "an event" announced by a sender whereas a reporter says, "an opposite event". The RSU then collects feedback from the vehicles (trusted clarifiers) which are visiting the presumed event location thereafter. With this data, the RSU decides, and sends rewards and punishment to the respective drivers when the decision is ready. In comparison to TPD reward and punishment, the RSU reward and punishment mechanism is simple, and it only assigns a fixed reward or punishment to the disputed drivers irrespective of the severity of incident, driver past behaviour, and RSU confidence in the sender or reporter (environmental dynamics). Thus, in this paper an advanced RSU reward and punishment generator is developed to assign a justified level of reward or punishment to the drivers concerned. It is found in the state-of-the-art that many researchers use Mamdani fuzzy logic to deal with the imprecision and uncertainty. In the Mamdani fuzzy logic, each rule output is a fuzzy set, and the rules can be designed intuitively with some expert knowledge.

In [6], an RSU assigns only fixed rewards and punishment based on the decision and does not consider the environmental dynamics. It is found that there is some uncertainty and incompleteness involved in the dispute resolution process. Thus, our attention is drawn to developing an advanced model to reward or punish drivers using a fuzzy logic based RSU assessment method. This model considers various factors and then allocates justified levels of reward or punishment accordingly. Also, in [6] the driver behaviour is modelled with a straightforward probabilistic distribution. Drivers with higher trust scores send less untrue messages. However, the probability distribution is fixed and not influenced by the changing trust score of the driver. This is why a Markov-chain based driver behaviour model is introduced. The states of the Markov model are associated with a specific range of trust scores. From each state, a driver's lying probability is defined which controls their likelihood of making trustworthy or malicious announcements. From a higher trust state, drivers announce less untrue messages whereas from the lower trust state they are more likely to announce untrue messages. The following contributions are made in this paper:

- The RSU reward and punishment mechanism is amended using Mamdani fuzzy logic-based assessment. This method considers the severity of incident, confidence score in the sender or reporter and driver past behaviour.

- A Markov-chain based driver behaviour model is used to govern the behaviour of drivers from different trust states. The state transition model along with the conditions to move between states is given. Also, from each state, the trustworthy / malicious announcement probability is defined.

- A series of experiments have been conducted to validate and compare the performance of the fuzzy versus fixed reward and punishment schemes.

- The Markov chain behaviour model is examined by defining the probabilistic distribution of sender, reporter drivers from different trust states and changing the initial trust distribution of drivers.

The paper is organized as follows: Section II reviews trust models based on fuzzy logic and Markov chain-based models and similar state-of-the-art. Section III briefly introduces author's earlier trust framework and presents the proposed fuzzy logic based RSU reward and punishment assessment method as well as Markov Chain driver behaviour model. Section IV describes the simulation environment and parameters for the experiments. Section V gives analysis and validation of results. Section VI compares fuzzy versus fixed RSU rewards and punishments and analyses the driver behaviour model with changing trust scores. Section VII presents the discussions. Finally, this work is concluded in the Section VIII where possible future research directions are indicated.

## II. Literature Review

This work primarily implements a fuzzy logic-based reward and punishment mechanism at RSUs. Additionally, a Markov model-based driver behaviour is developed to control the behaviour of drivers. These are improvements to the trust framework presented in [6]. First, some of the existing state-of-the-art trust models are briefly reviewed including fuzzy logic and Markov-model approaches. Trust approaches vary from different perspectives. For example, they can be differentiated based on whether they are application-oriented (architecture-less) [14] or architecture-based [3, 4]. Some approaches are centralized like [15] whilst others employ a decentralized architecture like [16]. Also, they can differ based on their data collection mechanism. For example, some schemes use only direct recommendations as [17]; others use both direct and indirect recommendations like [12, 34] for trust evaluation. Trust evaluation mechanisms are divided into three main classes which are Entity-Oriented Trust Models (EOTMs), Data-Oriented Trust Models (DOTMs), and Hybrid Trust Model (HTMs). These trust evaluation mechanisms are briefly reviewed next.

### A. Entity-Oriented Trust Models (EOTMs)

Entity-oriented trust models are epitomized by [10], where the researchers securely manage allocated credit using a Tamper Proof Module (TPM) on every vehicle. A vehicle first gets the transmission cost and the signed message from its TPM. Receiver vehicles consider the sender's reputation to decide whether to trust a message and the trust is revised using feedback from all receivers. This approach considers the presence of false attacks and benevolent vehicles. However, the process for setting a revised trust score can lead to excessive communication. In [18], the researchers consider familiarity, packet delivery ratio, timeliness, and interaction frequency to manipulate a weight-based aggregated final trust. They analyse the time-aware trust of vehicles from histories of interactions. However, they do not consider any attacker model for validation. In [4], a trust model uses a false message detection scheme to generate feedback on received messages which computes the trust of vehicles. Vehicles utilize primary and secondary scores from the RSUs for further communication until the next periodic update. The scheme is evaluated in the presence of false messages for both urban and highway environments. Nevertheless, it suffers from excessive trust metric dissemination.

In [15], a Reputation-based Global Trust Management (RGTE) scheme employing a Reputation Management Center (RMC) is presented. The RMC keeps track of the updated reputation of all vehicles. Every vehicle sends its recommendation about its neighbours to the RMC and then it uses central limit theory to exclude unreasonable recommendations. It updates reputation of vehicles for which it receives recommendations. Whenever a receiver receives a message, it directly consults the RMC about the reputation of the sender. However, in this model, the server is contacted frequently for reputation requests and replies.

A fuzzy logic-based direct trust and Q-learning-based indirect trust is considered in [12]. This approach analyses precision and recall metrics with varying the number of malicious vehicles. However, the overhead is high as it involves repeated sensing of messages from neighbours. The authors in [19] apply fuzzy logic to calculate trust using experience, plausibility, and location accuracy. Furthermore, location accuracy is determined using fog nodes. It can detect bogus attacks and message alteration attacks. However, vehicles consulting with fog nodes for location accuracy raise the communication overhead. The authors in [20] also use fuzzy logic and calculate the relaying trust and coordinating trust. Then the final trust is computed from these two and a path is identified using a set of rules and experiences. However, this model only considers trust-based routing to deliver a message along the most trusted path.

The study [21] selects an optimal path for packet forwarding using fuzzy logic-based transmission method. In this method, driving direction, vehicle speed, link time, hop count are used for relay node selection. Additionally, it considers the future state of vehicles. In [22], a fuzzy logic-based trust model is proposed that uses the RSU assessment, emulation attack attempts, and collaboration degree to assess the trust of vehicles. It incentivises good behaviour and punishes malicious vehicles. However, their analysis only concentrates on network performance measurement considering the malicious behaviour of making the connection slow, modifying messages, and stating false opinions.

In [23], a fuzzy logic-based trust model is proposed to address uncertainty and inaccurate trust estimation in a VANET. In this method, edge servers compute the trust of vehicles using fuzzy logic from packet drop, alteration, and false message injection factors. The analysis considers message alteration attacks and bad-mouthing attacks. In [24], a fuzzy logic-based system is used for vehicle authentication. This system only considers distance and trust factors to classify vehicles as partially or fully trusted or malicious. However, this approach is not analysed in the presence of a known adversary.

In [25], a fuzzy-logic-based trust model is presented where plausibility, experience, and vehicle type are used to decide on the validity of events. The fuzzy decision-making module of receiver vehicles utilizes these factors to compute the trust of the sender to determine whether to accept or reject or to forward a message. The analysis considers simple, opinion tampering, and on-off attacks. Every receiver vehicle applies fuzzy logic independently to forward an announcement to a further vehicle. The researchers in [26] propose a Hidden Markov Model (HMM) based trust evaluation method which computes trust of vehicles at the RSUs. This model improves the accuracy in detecting malicious vehicles compared to a baseline scheme.

### B. Data-Oriented Trust Models (DOTMs)

In [27], the researchers present machine learning based trust models (i.e. KNN, decision tree, naïve Bayes, and random forest). An RSU runs a location spoofing attack detection framework which uses stored data and received Basic Safety Messages (BSMs). The model is trained with both legitimate and malicious data. The analysis examines the accuracy, precision, and recall for all machine learning approaches. However, the analysis is limited to the BSM data. Research [28] differentiates malicious vehicles from benevolent ones using an ensemble learning algorithm and a decision tree-based model. The analysis includes measuring the accuracy, precision, and recall. However, it only identifies fake positional data.

In [29] author proposes a fuzzy system considering network density, relaying distance, and trust inconsistency to predict the relaying trust of vehicles. Then the coordinated trust is computed using velocity, connection degree and loss parameters. After that, the final trust is computed using a fuzzy system considering the relaying and coordinated trust that is used to find a trusted path. However, the model only selects the trusted relayer to confirm the trusted path for delivering messages. In [30], the authors propose a data oriented HMM-based reputation model. This model evaluates the reliability and the legitimacy of the announced messages. The reputation of vehicles is updated based on the correctness of safety and non-safety messages. The study [31] presents a vehicle behavioural monitoring and trust computational model to classify fake and legitimate messages. This model uses a neuro-fuzzy method to evaluate the behaviour of vehicles. It features accurate malicious message detection from speed and emission data. Using this data, the model can isolate misbehaving vehicles and discard messages from them.

### C. Hybrid Trust Models (HTMs)

In [32], the researchers present a Markov Chain-based hybrid trust model for VANETs. In this scheme, a state transition model, and the state transition probabilities are presented considering a cooperation factor and the accurate evaluation of messages. The monitoring process considers trustworthy message broadcasting besides cooperativeness, and they examined camouflaged behaviour. The researchers in [33] consider the likelihood and impact of taking a decision when both the event and the opposite event coexist. This approach is compared with a multi-faceted trust model. The results suggest that this approach always selects a low-risk action relative to a typical trust-based approach. However, the model is designed for a clustered environment.

In [34], the researchers develop a Bayesian inference-based direct and recommendation-based trust model. The direct trust considers penalties and time-decaying information. Also, the confidence of direct trust is checked beforehand to avoid unnecessary recommendation trust calculations. The analysis considers packet drop and interception as malicious behaviours. Alternatively, in [35], a self-organizing hybrid trust model is proposed for both urban and rural scenarios. This approach keeps a history of interactions and then validates the received messages by assigning a credit. This model accepts the message with the highest trust for a particular event. It can detect fake event locations, source locations, and event times as well as revoke messages from malicious vehicles. However, this model is not evaluated against a baseline. Study [36] embeds the trust certificate of a vehicle with the message that a receiver uses as a weight to evaluate the trust of the data. A vehicle that visits the event location either confirms or denies the event. The vehicle sends all stored feedback to an RSU to forward it to the Certificate Authority (CA) to update a vehicle's trust certificate. Later, vehicles receive updated trust certificates from the CA via an RSU. Thus, the approach suffers from communication overhead to frequently update trust certificates.

In [37], trust is computed from past experiences, neighbouring vehicle information, trust of the vehicle, and the packet delivery ratio. This approach has a trust manager, route manager, and decision manager. The trust manager finds the path trust and calculates the required time to forward a message to the destination. The decision manager informs a nearby RSU if the vehicle does not want to participate in packet forwarding. This model selects a path with the highest trust and lowest delay. The approach considers the packet delivery ratio, delay, and the number of routes. However, they only implement the trusted routing. In [38], a vehicle learns cognitively from the environment and develops contexts around an event to infer the trust. It defines a context which associates a set of interrelated concepts (for example vehicle, evaluation, event). This framework considers experience, opinion, and role for the trust evaluation. For outlier detection, time, speed, and distance thresholds are used. Besides finding the trust level for every report, this approach also finds the confidence of the report. The framework is simulated in both rural and urban scenarios and compared against existing frameworks. However, malicious vehicles can bypass the outlier-based detection process and can send false messages within the acceptable threshold they set for this model. In [39], an RSU is solely responsible for the trust computation of vehicles, and it collects recommendations and feedback from vehicles. Besides this, the RSU creates, manages, and merges clusters for the VANET. The scheme is robust against thwarting Sybil and wormhole attacks. The RSU also identifies malicious vehicles and prevents them joining another cluster. Though they maintain trustworthy clusters, this requires considerable dissemination and cluster management at the RSUs which demands significant computational effort.

III. Proposed RSU Assessment Method and Driver Behaviour Model

The proposed RSU assessment method is used only to assign RSU reward and punishment to drivers who are involved in disputes relating to untrue attack dissemination in the network. This is an extension to the trust framework described in [6]. The Markov model is used for behavioural analysis of drivers with this trust framework.

*A. Sender – Side Trust Framework*

In [6], a trust management framework is presented where a Tamper Proof Device (TPD) is fitted to each regular vehicle providing trust-based access control to the VANET. This framework considers regular vehicles, along with police, ambulance, and fire service vehicles. The main components of the framework are the vehicles, RSUs, and the Trust Authority (TA). RSUs send incident data to the TA for storage. Vehicles take different roles based on their activities in the network. When a vehicle announces a message, then it is a sender vehicle. When a vehicle receives a message, it is called a receiver vehicle. When a vehicle notices an announcement is invalid, it can become an untrue attack reporter. However, this report can be malicious as well for which the framework arranges some punishment upon an RSU ruling. An RSU collaborates with the vehicles which are visiting the event location near the time to decide on the validity of the event. The vehicles which send feedback when collaboration is running are called clarifiers.

The following equations define the trust thresholds to achieve access control. Equation (1) confirms the trust score of a driver stays in the range of 0.05 to 0.9 irrespective of trust adjustments. Equation (2) relates to access-blocking of a driver/vehicle. Equation (3) regulates the message relaying ability and Equation (4) determines ability of regular vehicles to make announcements.

$$T_i = \begin{cases} 0.05, & T_i < 0.05 \\ 0.9, & T_i > 0.9 \\ T_i, & 0.05 < T_i \leq 0.9 \end{cases} \quad (1)$$

$$T_i = \{Blacklist. \; T \leq 0.05\} \quad (2)$$

$$Message\ Relaying\ Ability = \begin{cases} False, & 0.05 \leq T < 0.25 \\ True, & T \geq 0.25 \end{cases} \quad (3)$$

$$Message\ Generation\ Ability = \begin{cases} Limited, & 0.05 \leq T < 0.5 \\ All, & T \geq 0.5 \end{cases} \quad (4)$$

Within this framework, regular vehicles are classified as access-*blocked* (T = 0.05), *not trusted* (0.05 < T ≤ 0.25), *lowly trusted* (0.25 < T < 0.5), *trusted* (0.5 ≤ T < 0.8), and *highly trusted* (0.8<T≤0.9). The trust of official vehicles is T = 1.0 which is higher than the maximum trust of a regular vehicle (T=0.9). A set of rules are employed for governing the actions of regular vehicles [6].

*1) Trust based access control for message announcements:* It is assumed each driver can announce an event if it is seen in the dashboard. It is dynamically updated based on the driver's trust score. Messages are organized into classes and each class is associated with a range of trust scores for access control. Vehicles must achieve a particular trust score to announce messages of a certain class. The framework rewards trustworthy announcements from the TPD after expiry of a withhold timer and optionally penalizes a driver if a driver delays beyond an acceptable limit. The TPD updates trust in relation to announcements, reporting, clarifying, relaying, and beaconing besides adjusting trust with RSU rewards and punishments.

*2) Functional diagram of the framework:* Assume, a trusted sender sends an announcement based on what he/she observes on a road which receivers receive and relay. The event is reported (opposite event) by a reporter after he/she thinks that the event has not occurred at the said location. An RSU upon reception of the report starts collaboration to decide on the truthfulness of the event. Then it informs the TA of its decision and sends fixed rewards (0.1) / punishments (0.1) to the respective drivers based on the decision. The TPD of the respective drivers combines the RSU assessment with the driver's current trust. Additionally, the TA decides on whether blacklisting of a driver is necessary. This decision is conveyed via an RSU and the driver's TPD implements it. The functionality is shown in Fig. 1.

*3) RSU untrue message detection:* When a reporter reports an untrue attack, an RSU resolves the issue using a sum of weighted feedback calculation. This feedback data are collected from clarifier vehicles. However, when an official vehicle sends feedback, an RSU directly uses this to decide on the dispute. The TA also maintains a driver profile database consisting of the recent records from disputed decisions. A decision results in either a reward or punishment for a driver which are saved into this list. The untrue message detection mechanism is shown in Fig. 2.

In [6], the untrue detection process executes at RSUs to allocate the fixed reward and punishment without considering the severity of incident, driver past behaviour, and RSU confidence in the sender or reporter. Thus, the reward and punishment scheme lacks a sophisticated model to assess the appropriate magnitude of the reward or punishment. These parameters are important to consider as they are related to the event and the driver. They also vary from one event to another, from one driver to another, and the collected feedback. A fuzzy logic-based reward and punishment scheme is a good fit as these parameters are uncertain and inexact, although the reward or punishment should be based on the severity level of the incident. Also, fuzzy logic can approximately imitate the human-level decision making. The fuzzy logic based RSU controller can account for various factors and assign a justified level of reward or punishment for a given driver.

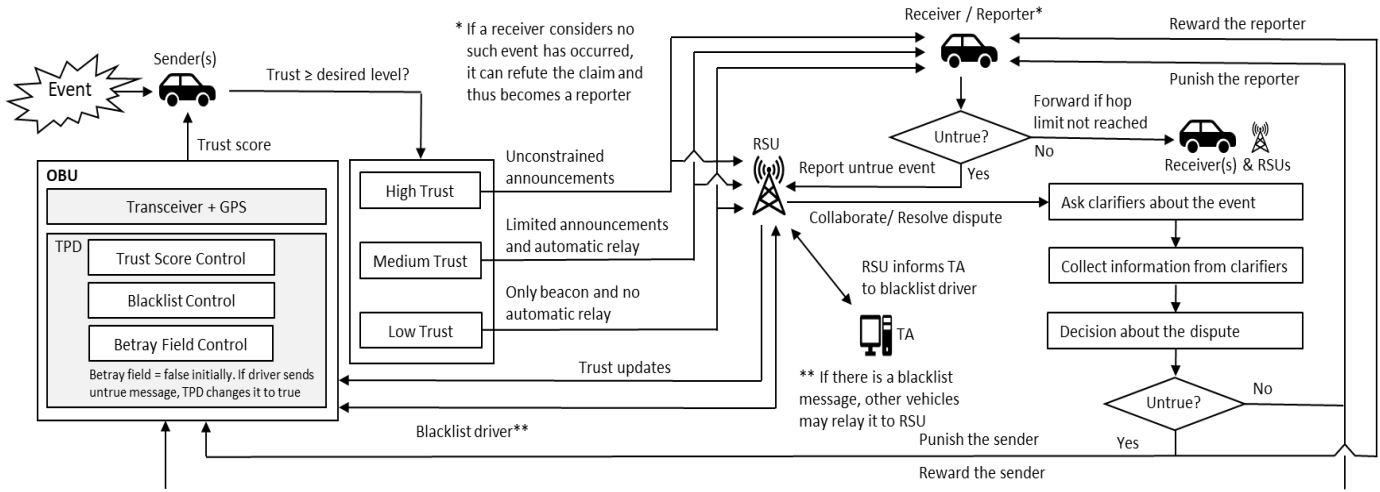

Fig. 1. Functional diagram of the trust framework [6].

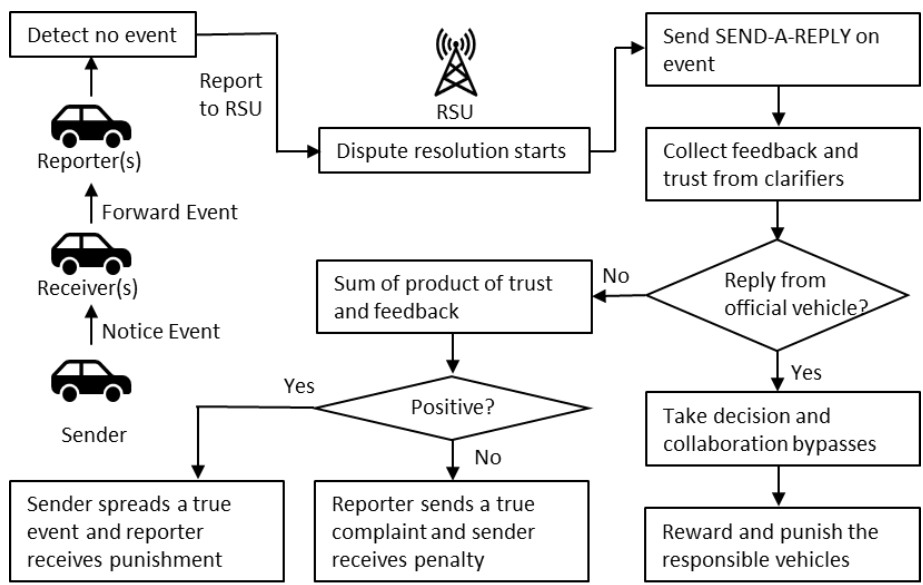

Fig. 2. RSU steps for untrue message detection.

*B. Overview of the Proposed Fuzzy RSU Reward or Punishment Assessment Scheme*

Fig. 3 depicts the proposed fuzzy RSU controller for determining reward or punishment. It starts from the left-hand side where it collects three inputs which are driver past behaviour, confidence in the sender or reporter and severity of the incident. This involves some form of pre-processing of input data to feed into the fuzzy controller. Then these inputs are handed over to the fuzzifier to produce input fuzzy sets. These sets are delivered to the fuzzy inference module which evaluates the fuzzy rules on the input fuzzy sets to produce the output fuzzy sets. These sets are then transferred to the defuzzifier module to generate the crisp number as output variables which is sent to the respective drivers as the level of reward or punishment for their action. A disputed decision at an RSU invokes the execution of this function to calculate the extent of reward or punishment for a conflicted announcement.

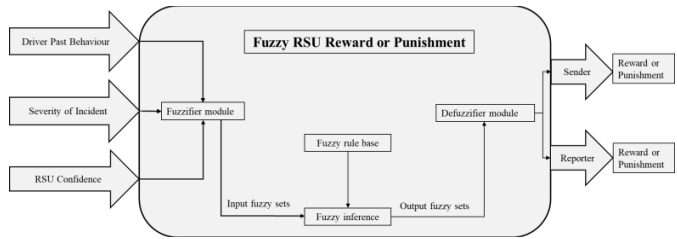

Fig. 3. A block diagram of the proposed fuzzy logic based RSU controller.

*1) Fuzzification:* Fuzzification finds the degree of membership for each input to the fuzzy sets (one or more linguistic variables) using a membership function. To find the degree of belonging, first the shape of the membership functions for every input are defined. Then the degree of belonging to the fuzzy sets are determined for each input. Membership functions are defined intuitively with the help of linguistic variables as shown in Table I.

TABLE I. INPUT FUZZY SETS

| Input Parameters | Fuzzy sets |
|---|---|
| Driver Past Behaviour (DPB) | Good (G), Neutral (N), and Bad (B) |
| Severity of the Incident (SI) | Not Severe (NS), Less Severe (LS), and High Severe (HS) |
| RSU Confidence (RCS) | Low (L), Medium (M), and High (H) |
| Reward/ Punishment | Very Low (VL), Low (L), Medium (M), High (H), and Very High (VH) |

*a) Driver Past Behaviour (DPB):* The RSU uses a membership function to convert each input to the degree of belonging to the fuzzy sets. RSUs always send data concerning the rewarded and punished drivers to the TA. An RSU asks for DPB data from the TA. Let, NoP and NoR be the recorded number of rewards and punishments for the concerned drivers from their previous disputed events. When the TA sends NoP and NoR data to a dispute resolver RSU, then it estimates the ratio of NoP/(NoR+ NoP) for the relevant drivers. The RSU feeds this data directly into the fuzzifier to get a degree of belonging of the DPB to each from the set:

{"Good", "Medium", "Bad"}. The DPB ranges from 0 to 1 and each DPB value is separated by 0.1. For example, if a driver record contains 4 punishments out of the 10 most recent records, then the DPB is 0.4. The fuzzification returns the fuzzy value as {Good: 0.24, Medium: 0.76, Bad:0}. Fig. 4 shows the membership function for driver past behaviour.

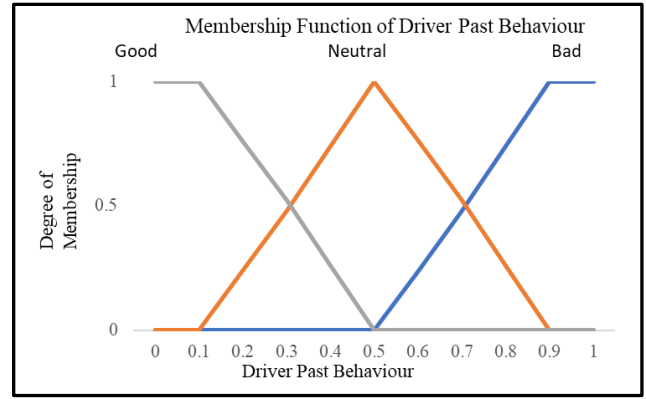

Fig. 4. Membership function for driver past behaviour.

*b) Severity of Incident (SI):* The list of potential events is shown in Table II for this fuzzy controller. This is just an example list of possible events. In this Table, the event's name, and its severity (assumed impact on human lives) are shown.

TABLE II. POSSIBLE EVENT LIST

| Incident Name | SEVERITY LEVEL (LOWEST TO HIGHEST) |
|---|---|
| Road Clear | 0 |
| Debris or Road Spillage (Oil or Muds or Sands) | 1 |
| Illegal Waste Dumping | 2 |
| Poor Conditioned Road | 3 |
| Minor Road Defect (Faded Sign) or Malfunctioning Traffic Element | 4 |
| Stranded or Abandoned Vehicle or Obstacle or No Obstacle | 5 |
| Major Road Defect (Pothole, Illegal Sign) | 6 |
| Diversion or Road Maintenance | 7 |
| Severe Weather (Snowy Road or Poor Visibility Due to Fog etc) or Environmental Incident | 8 |
| Flood or Fallen Tree on Road | 9 |
| Congestion | 10 |
| Traffic jam | 11 |
| Accident | 12 |

Every RSU stores a copy of this table. When there is a dispute, the RSU looks up the severity level from the table to feed into the fuzzifier. Three fuzzy sets {"Not Severe", "Less Severe", and "High Severe"} are used for this input. When the SI is inputted, the fuzzification returns the fuzzy value as {Not Severe: 0.18, Less Severe: 0.82, High Severe: 0}. Fig. 5 shows the membership function for the severity of incident.

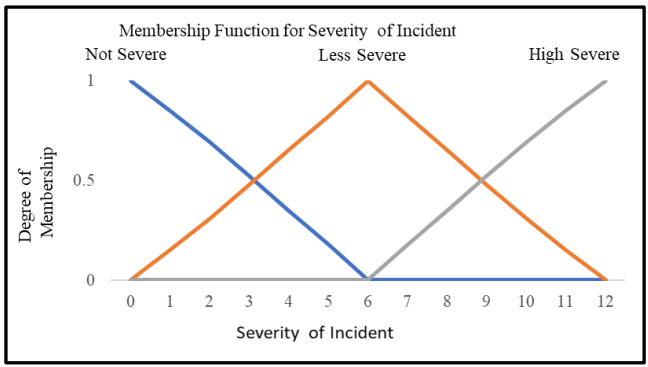

Fig. 5. Membership function for severity of incident.

*c) RSU Confidence in the Sender or Reporter:* An RSU obtains a confidence score from the received feedback using the ratio of feedback that supports the sender's event to the sum of feedback which supports and contradicts the announcement. This is the RSU's confidence in the sender. Similarly, the RSU confidence in the reporter is defined as the ratio of feedback that supports the reporter's report to the sum of the feedback which both supports and contradicts the reporter's report. Three fuzzy sets are defined for the RSU confidence which are {"Low", "Medium", and "High"}. The RSU confidence in the sender or reporter may or may not differ based on the feedback. The fuzzification returns the fuzzy value as {Low: 0, Medium: 0.33, High:0.67} for RSU confidence. Fig. 6 shows the membership function for RSU confidence in the sender or reporter.

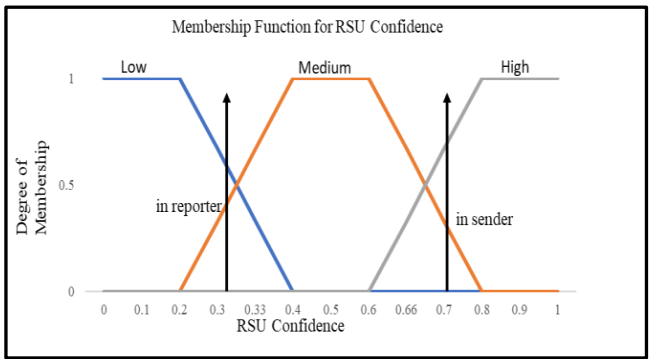

Fig. 6. Membership function for RSU confidence.

*2) Fuzzy rules for reward and punishment:* There is a separate set of rules for reward and punishment. Table III shows the set of rules used for rewarding whereas Table IV is used for punishing drivers. The reason for maintaining two sets of rules is that for one situation the reward may be smaller, but the punishment should be higher. Let output membership be OM. As each input has three fuzzy sets, thus the total number of rules is 3*3*3=27 which are given next. The first rule from Table III says as "if the (Driver Past Behaviour (DPB) is Good) AND (Severity of Incident (SI) is Not Severe (NS)) AND (RSU Confidence (RC) is Low), then the Reward is Low". This explanation goes to other rules as well.

TABLE III. FUZZY RULES USED FOR REWARD

| Rules | DPB | SI | RC | R |
|---|---|---|---|---|
| 1 | Good | Not Severe | Low | Low |
| 2 | Good | Not Severe | Medium | Medium |
| 3 | Good | Not Severe | High | High |
| 4 | Good | Low Severe | Low | Medium |
| 5 | Good | Low Severe | Medium | High |
| 6 | Good | Low Severe | High | Very High |
| 7 | Good | High Severe | Low | High |
| 8 | Good | High Severe | Medium | Very High |
| 9 | Good | High Severe | High | Very High |
| 10 | Neutral | Not Severe | Low | Low |
| 11 | Neutral | Not Severe | Medium | Low |
| 12 | Neutral | Not Severe | High | Medium |
| 13 | Neutral | Low Severe | Low | Low |
| 14 | Neutral | Low Severe | Medium | Medium |
| 15 | Neutral | Low Severe | High | High |
| 16 | Neutral | High Severe | Low | Medium |
| 17 | Neutral | High Severe | Medium | High |
| 18 | Neutral | High Severe | High | Very High |
| 19 | Bad | Not Severe | Low | Very Low |
| 20 | Bad | Not Severe | Medium | Very Low |
| 21 | Bad | Not Severe | High | Low |
| 22 | Bad | Low Severe | Low | Very Low |
| 23 | Bad | Low Severe | Medium | Low |
| 24 | Bad | Low Severe | High | Medium |
| 25 | Bad | High Severe | Low | Low |
| 26 | Bad | High Severe | Medium | Medium |
| 27 | Bad | High Severe | High | High |

TABLE IV. FUZZY RULES USED FOR PUNISHMENT

| Rules | DPB | SI | RC | P |
|---|---|---|---|---|
| 1 | Good | Not Severe | Low | Very Low |
| 2 | Good | Not Severe | Medium | Very Low |
| 3 | Good | Not Severe | High | Low |
| 4 | Good | Low Severe | Low | Low |
| 5 | Good | Low Severe | Medium | Low |
| 6 | Good | Low Severe | High | Medium |
| 7 | Good | High Severe | Low | Medium |
| 8 | Good | High Severe | Medium | High |
| 9 | Good | High Severe | High | High |
| 10 | Neutral | Not Severe | Low | Low |
| 11 | Neutral | Not Severe | Medium | Low |
| 12 | Neutral | Not Severe | High | Low |
| 13 | Neutral | Low Severe | Low | Low |
| 14 | Neutral | Low Severe | Medium | Medium |
| 15 | Neutral | Low Severe | High | Medium |
| 16 | Neutral | High Severe | Low | Medium |
| 17 | Neutral | High Severe | Medium | High |
| 18 | Neutral | High Severe | High | Very High |
| 19 | Bad | Not Severe | Low | Very Low |
| 20 | Bad | Not Severe | Medium | Low |
| 21 | Bad | Not Severe | High | Low |
| 22 | Bad | Low Severe | Low | Low |
| 23 | Bad | Low Severe | Medium | Medium |
| 24 | Bad | Low Severe | High | High |
| 25 | Bad | High Severe | Low | Very High |
| 26 | Bad | High Severe | Medium | Very High |
| 27 | Bad | High Severe | High | Very High |

*3) Fuzzy inference:* Human decision making can be approximated by using fuzzy inference. Fuzzy Inference produces fuzzy output sets from the input fuzzy sets. During the fuzzy inference, each rule executes sequentially to obtain the desired output fuzzy set. A rule executes when its antecedent is satisfied. The antecedent of each rule is formed using Fuzzy AND, Fuzzy OR and Fuzzy NOT. The Fuzzy AND and Fuzzy OR are used as fuzzy logical operators. Fuzzy AND returns the minimum of all membership values from the antecedent part whereas a Fuzzy OR returns the maximum to clip or bound the height of output membership function. The returned value from each rule is the firing strength which is used to clip or bound the height of the output membership function. This means the output of the antecedent define the corresponding degree of membership value of the consequent part of each rule. Fig. 7 shows the output membership function of the reward/punishment where the reward is 0.08 and the punishment is 0.03.

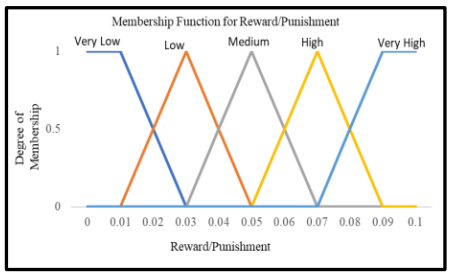

Fig. 7. Output membership functions for reward and punishment.

*4) Aggregation:* In this step, all outputs from the fuzzy inference are combined to get one aggregated fuzzy output set which is fed into the defuzzifier module to get the fuzzy reward and punishment. During aggregation, all similar output fuzzy sets are merged into one and their resultant fuzzy set has the maximum consequent from all similar output fuzzy sets. For example, if three rules produce **Low** output fuzzy set with the degree of membership are 0.05, 0.064, and 0.021, then the aggregation combines these into one **Low** output fuzzy set with the degree of membership equals 0.064.

*5) Defuzzification:* A defuzzification method takes the aggregated output fuzzy membership function and produces one crisp number which is the desired output from this system. Centre of Gravity (COG) is the most widely accepted defuzzification method to find the final defuzzified value. It is the final step of the fuzzy system. The most widely defuzzification method of Mamdani inference is the centroid technique. It delivers a point where a vertical line divides the aggregated output fuzzy set into two equal masses. This method finds a point which represents the COG of a fuzzy set, A, on the interval [a, b]. Here, $\mu$ denotes the degree of membership. A reasonable estimation can be obtained by sampling a set of points. This is expressed as in Equation (5).

$$CoG = \frac{\int_b^a \mu_A(x)x\,dx}{\int_b^a \mu_A(x)\,dx} \qquad (5)$$

*6) An Example Fuzzy Inference for reward*

*a) Fuzzy Inference:* In Fig. 8, the truncated execution of two rules for calculating fuzzy reward is shown as they are selected during the fuzzy inference. The execution of other rules is deleted deliberately to save space. The antecedent part of the rules is evaluated first to generate an output from each rule with the height defined by the min or Fuzzy AND operation of the antecedents. The DPB is 0.8, SI is 4, and the RSU confidence is 0.33 for example fuzzy inference. Similarly, fuzzy punishment is determined using the rules from Table IV.

*b) Redundant rule reduction for reward:* When multiple rules produce the same output fuzzy set with different values, they can be combined into one by taking the maximum of all consequent values for the same output fuzzy set; As the rules 10, 11, 13, and 23 have Low output fuzzy set, so taking the maximum gives us Rule 23 with 0.65 as the membership degree for the Low output fuzzy set. As there is only one Medium fuzzy set, it is included directly. Also, Rules 19, 20, and 22 have the Very Low output fuzzy set, thus the maximum consequent value from these three rules is Rule 20 to include in the selected group for aggregation. This situation is depicted in Fig. 9.

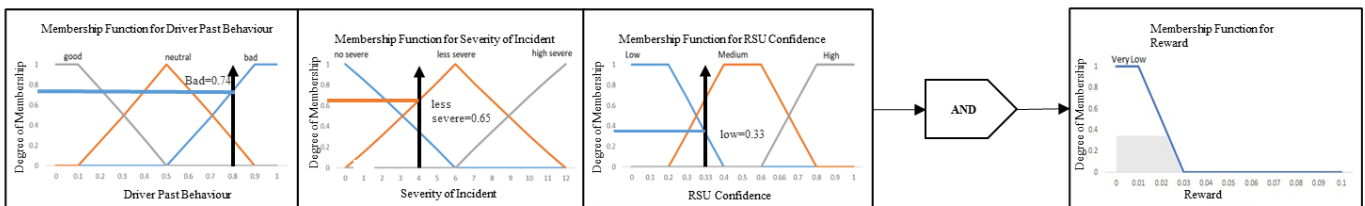
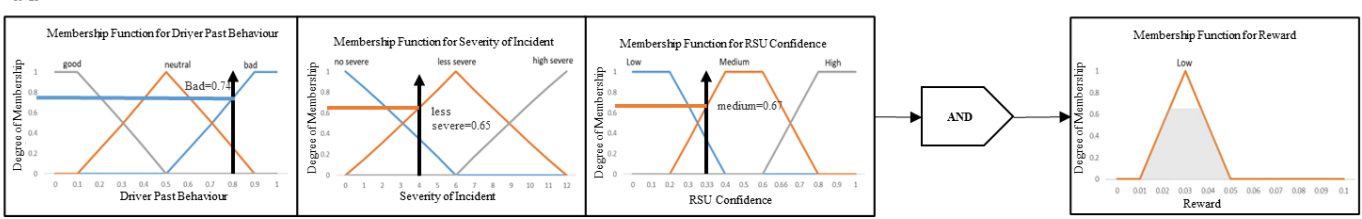

Fig. 8. Fuzzy rule inference for reward assessment.

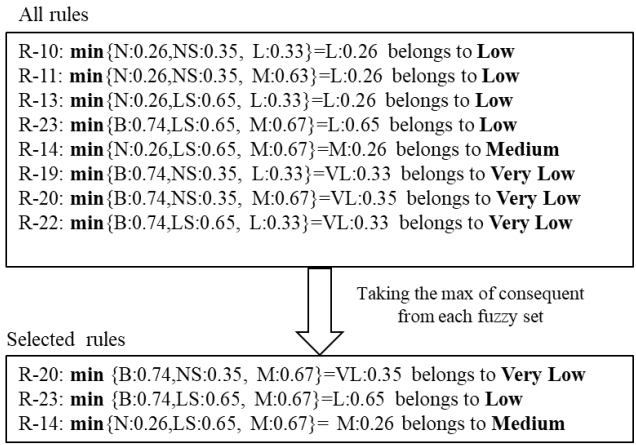

Fig. 9. Redundant rules reduction for reward.

*c) Aggregation of the consequents for reward:* The aggregation is applied to the selected rules which merges them to get the combined output membership function. In this step, only the output fuzzy sets with the highest degree of membership are used where all the output fuzzy sets with a lower value are inclusively covered. This is a combined fuzzy set as depicted in Fig. 10.

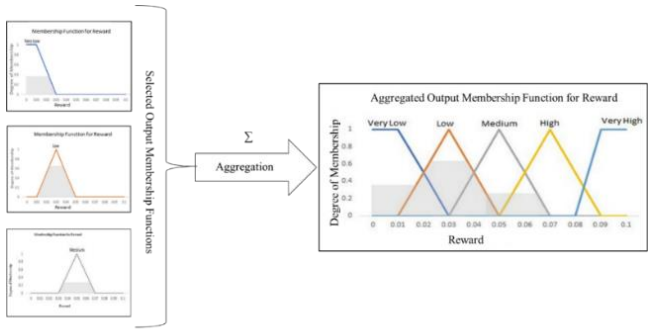

Fig. 10. Aggregated output membership for reward assessment.

*d) Defuzzification for reward:* Fig. 11 shows the assessed reward from the centroid defuzzification method. First, the area is sliced equally as shown in Fig. 11. Then the reward of 0.030014 is obtained as shown with a green arrow on the x-axis.

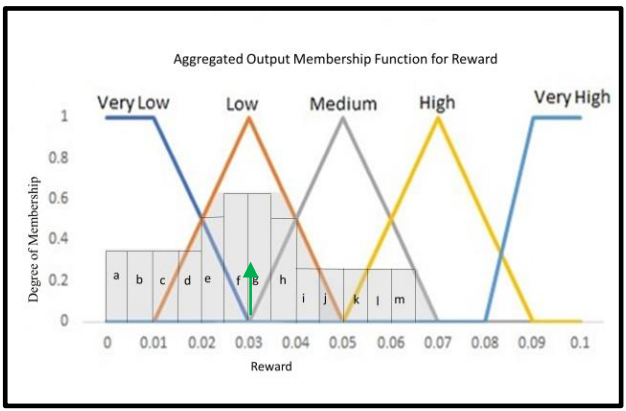

Fig. 11. Defuzzified reward for the example case.

## C. Markov Chain Driver Behaviour Model

Driver announcements are only randomly reported by some reporters with the attack generation probability is set to 0.4 in the analysis of the current model [6]. In the series of experiments, a driver's behaviour is not modelled to see at what situation they are sending more trustworthy or malicious messages. The disputes only arise from the reporter's untrue attack messages which are generated randomly when the probability function returns true. This is why a model is developed which can control the message announcement behaviour from the driver. To this end, Markov-chain state transition model is created which can provide driver behaviour modelling and control message announcements. The proposed driver behaviour model is defined with some fixed states and from each states message announcement probability for both trustworthiness and maliciousness are defined. There are some fixed conditions to switch between the states of this model.

The trust states of the proposed Markov chain model are defined with drivers lying probabilities to examine their honesty or lying behaviour. These states are defined based on the different trust thresholds set for the framework. Trust states are ordered according to the increasing trust values. Thus, a driver who wants to reach a higher trust state must achieve a higher trust value by announcing only trustworthy messages. A driver switches to another state when its trust score falls outside the range of trust scores for the current state. It is believed that a driver with a higher trust state possess the higher probability to announce more trustworthy messages than those with a lower trust state. With this model, acceptable behaviour means announcing trustworthy messages whereas the unacceptable behaviour means announcing untrue messages. When a trustworthy message is announced, a driver improves the trust score from it. If another driver sends a report about it and the sender driver wins the dispute, then RSU reward is added with the current trust. As a result, the sender driver possibly makes a transition to another state which is associated with higher trust scores than the current one. In contrast, a driver loses trust score from the announcement of an untrue message when another reporter sends an untrue attack about it and the message is proved malicious by an RSU. Whether an announcement would be trustworthy or untrue, it is directly related to the driver behaviour. Hence, these activities are modelled with the proposed Markov chain-based state transition diagram by setting the probabilistic distribution to control untrue and trustworthy message announcements from each state. From each state, a driver earns rewards from the announcement, clarifying, reporting, forwarding and gets either reward or punishment from an RSU if there is a dispute relating to his/her announcement.

The proposed Markov model has six different trust states out of which one is the access-blocked state. A driver reaches this state when he/she is blacklisted, and his/her trust becomes 0.05. Other states are associated with different ranges of trust values. The six trust states are: "very good", "good", "normal", "bad", "very bad" and "access-blocked". The probabilities of sending trustworthy and untrue messages from these states are set as shown in Table V which can be configured with different values to simulate the variation in driver behaviour. Table VI lists the probability of sending untrue attacks in the different trust states which defines the behaviour of the reporter drivers. These values are selected such that drivers with higher trust states send less untrue messages and reports than in the lower trust states. In a real-world scenario, a driver can react differently at different times which can be modelled with a Markov chain-based driver behaviour model using a different probabilistic distribution.

With these trust states, a Markovian state transition-based driver behaviour model is presented, which is consistent with the trust framework described in Subsection III.A. A diagram of this model is shown in Fig. 12. It has fixed trust states, and each state is associated with a range of trust scores. A driver remains in a given state when his/her trust belongs to the range of trust values related to that state. With this model, a driver starts his/her journey from the "normal" state with a trust value equal to 0.5. From this state, a driver sends some announcements and relays events from others.

This model covers the announcement lying behaviour of drivers. Thus, from a "normal" state, a driver can build trust to reach the "good" state if he/she continues announcing trustworthy messages in the network. Also, he/she can lose trust by announcing untrue messages to reach the "bad" state from the "normal" state. He/she can even move to the "very bad" trust state if most of the announcements are untrue. In the worst case, the driver may be access-blocked if his/her trust score reaches 0.05.

Alternatively, from the "good" state, a driver can improve trust to move into the "very good" state to become a highly trusted vehicle. Once a driver is in the "very good" trust state, it is harder to lose trust as he/she only announces untrue messages with 0.1 probability. As such, the model captures the philosophy that good drivers tend to remain so, and vice versa unless they are encouraged to modify their behaviour. For consecutive untrue message announcements, a driver's trust score is reduced. In this case, he/she may be moved to the "good" or "normal" state. It is even possible to move into the "bad" or "very bad" state when he turns severely malicious. In this way, a mal-intent driver loses his/her trust and may be access-blocked in the network from where he/she cannot participate in any communication. When a vehicle is access-blocked, an external procedure is assumed to enable him/her to be reset to the "normal" trust state, if permitted.

TABLE V. DRIVER'S ANNOUNCEMENT LYING PROBABILITY

| Trust States | Probability of Announcing Trustworthy Message | Probability of Announcing Malicious Message |
|---|---|---|
| "very good" | 0.8 | 0.2 |
| "good" | 0.6 | 0.4 |
| "normal" | 0.4 | 0.6 |
| "bad" | 0.2 | 0.8 |
| "very bad" | 0.1 | 0.9 |
| "access-blocked" | 0 | 0 |

TABLE VI. REPORTER'S UNTRUE ATTACK REPORTING PROBABILITY

| Trust States | Probability of Reporting an Untrue Attack | Probability of Not Reporting an Untrue Attack |
|---|---|---|
| "very good" | 0.1 | 0.9 |
| "good" | 0.3 | 0.7 |
| "normal" | 0.5 | 0.5 |
| "bad" | 0.7 | 0.3 |
| "very bad" | 0.9 | 0.1 |
| "access-blocked" | 0 | 0 |

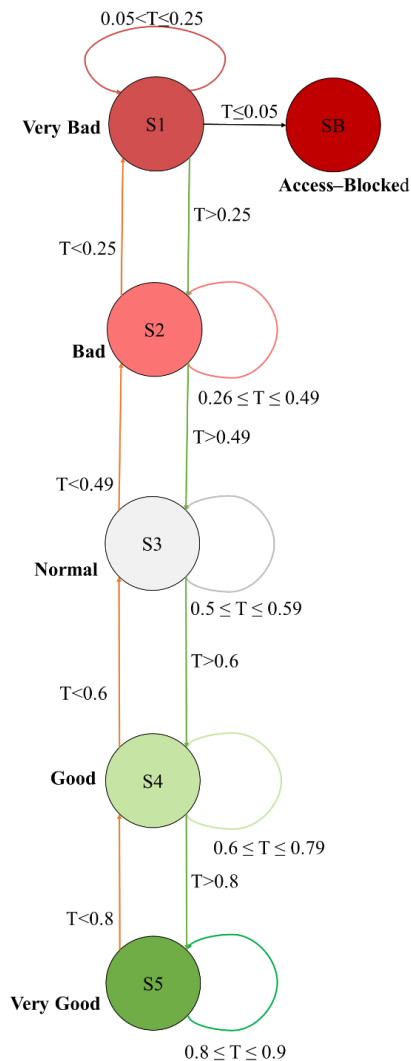

Fig. 12. Markov-chain behavioural model (state transition diagram).

A clarifier is a vehicle which sends feedback in response to an RSU query. This feedback is consistent with the driver behaviour model. This allows the behaviour of clarifiers to be programmed similarly to the probabilities defined for different trust states of the sender and/or reporter drivers. As the trust model does not evaluate a clarifier's feedback, their behaviour analysis is not considered as important as the sender or reporter information.

## IV. Implementation

First, Mamdani type fuzzy inference is implemented in MATLAB 2022 for RSU reward and punishment assessment. This provides a built-in fuzzy logic designer app where three inputs are created, their input membership functions, and corresponding fuzzy sets. After that, two different set of rules are entered into the rule editor of, and all the rules are given equal weight. As this is a two-output fuzzy system, two output membership functions and corresponding fuzzy sets are created. The fuzzy OR operator is used for punishment and the fuzzy AND is applied to the reward assessment. There are three fuzzy sets for each input and five fuzzy sets for each output. During the fuzzy inference for each dispute, all twenty sevens rules are evaluated individually to produce the fuzzy output sets for each output. The aggregation applies on these output fuzzy sets and then the centroid method is applied on the combined fuzzy sets to return the desired fuzzy reward and punishment. These output values from MATLAB are directly processed and inserted into two different lists in the OMNeT++ to be used with the proposed model. There are eleven possible values of DBP. Hence, for each DBP value, all possible values of SI and RSU confidence are considered. In this way, different combinations of input values are used with the fuzzy system. For each DPB, a list of values is produced, and a different data structure in OMNeT++ is created to enable faster searching for different combinations of input values.

When a dispute decision is ready, an RSU asks for the DPB data from the TA. As the TA maintains a list of past records for all drivers, it can serve the query readily. After that, the RSU calculates the DPB for the relevant drivers. The RSU also calculates a confidence score of the disputing drivers from the collected feedback. Additionally, the RSU determines the severity level of event. The RSU then looks up the corresponding fuzzy reward and fuzzy punishment from the list. These values are directly used in the reward and punishment messages which the RSU announces and forwards to nearby RSUs to announce, too. In this way, each respective driver/vehicle receives the fuzzy RSU reward and punishment.

The following set of experiments use the Markov chain-based driver behaviour model which is implemented inside the TPD of every regular vehicle. This model governs the driver's announcement behaviour by setting the probability of sending trustworthy and untrue messages based on the behaviour state.

## V. Analysis and Validation of the Markov Chain-Driver Behaviour Model

### A. Simulation Setup

A set of experiments has been carried out to evaluate the behaviour of sender or reporter drivers by changing their lying probability to observe the proportion of trustworthy and untrue messages generated from different trust states over the simulation period. The trust framework, the fuzzy reward and punishment mechanism, and the Markov state transition model are implemented in Veins [40] which comprises OMNeT++ [41] and SUMO [42]. It is an open-source framework which enables online communication between OMNeT++ and SUMO when the simulation is running. The participating vehicles run for 5000 simulation seconds (s) on a fixed circular route in the Erlangen city map shown in Fig. 13. 100 vehicles are added at the start of the simulation and their numbers are kept constant throughout the experiment. Vehicles undergo a warm-up period where they move without announcing any event. When the warm-up period has elapsed, a fixed sender driver announces messages periodically at 1000s periodic intervals for each event type starting from the 500s. The simulation includes multiple types of event announcement from the same driver of V[0] for behavioural analysis. The events are scheduled as an accident message at 500s, a debris message at 700s, a road defect message at 900s, a traffic element problem message at 1100s and a tree on the road message at 1300s. Reporters deterministically send untrue attack reports based on the probabilistic distribution defined in Table VI.

As it is required to model the behavioural change of these reporters as well, their trusts are shown in Fig. 15 to 20 beside the sender driver. In this way, a series of experiments are conducted with different initial trust distributions and then the trust evolution is observed to examine the distinctive driver behaviour. A fixed reward and punishment mechanism is used from the disputes to update the trust of drivers so the result can differentiate their behaviour, whether they lie or not and in what circumstances they lie. Other rewards and punishments within the trust framework are not enabled for this analysis of driver behaviour.

In this series of experiments, drivers can send untrue attacks even when their trust score is less than 0.5 which was not allowed with the trust model presented in [6]. If a driver can send a message from a particular trust state, then he/she is allowed to send an untrue attack version of the originated message. The RSUs employ a 120-second collaboration timer to determine the validity of a dispute from the clarifier feedback. Thus, the verification time delays the reception of rewards and punishments from an RSU. Also, RSU reward is disseminated in one message and RSU punishment is sent in another message to the driver which also adds an additional delay besides their availability to an RSU and wireless collisions. Thus, the collaboration timer and a vehicle's availability, delays the reception of reward or punishment at the vehicles concerned. Table VII lists the parameters for the experiments.

There are two sets of experiments conducted for examining driver behaviour model. In the first set of experiments, clarifiers send opinion based on the witness and a probability distribution. If a driver with a "very high" trust state generates an event, then the clarifiers send positive opinions with 0.8 probability and negative opinions with 0.2 probability. For the "good" trust state, clarifiers send positive opinions to 60% of cases and negative opinions to 40% of cases. A message from a "normal" trust state originating from a driver gets 40% positive and 60% negative opinions. From the "bad" state, clarifiers deny announcements 80% of the time and support only 20% of the time. From the "very bad" state, clarifiers deny announcements 90% of the time and support only 20% of the time. This distribution can be changed as needed to model the variation in a sender or reporter driver's behaviour. In the second set of experiments, clarifiers send feedback based on the probability distribution of their trust states as shown in Table VIII and the reporters send report based on the Table VI.

TABLE VII. SIMULATION PARAMETERS

| Parameters | Values |
|---|---|
| Fuzzy reward and punishment | Varied |
| Fixed reward and punishment | 0.1 |
| Data Collection Nature | 1. When all features enabled  2. When only RSU judgement applied |
| Simulation Period | 5000s |
| Warm-up Period | 500s |
| Announcement Interval | 1000s |
| Initial Trust | Uniform distribution (0.5-0.6) |
| Number of Vehicles | 100 |
| Multiple types of Events Generated | Yes |
| Number of RSUs | 12 |
| Number of TA | 1 |
| Attacker Model | Untrue and Inconsistent Behaviour |
| Untrue Attack Generation | Based on the message class |
| Announcement Reward | Maximum of 0.8 (0.1 to 0.8 based on delay and distance) |
| Clarifier Reward | Maximum of 0.8 |
| Relaying Reward | 0.002 |
| Collaboration Timer | 120s |

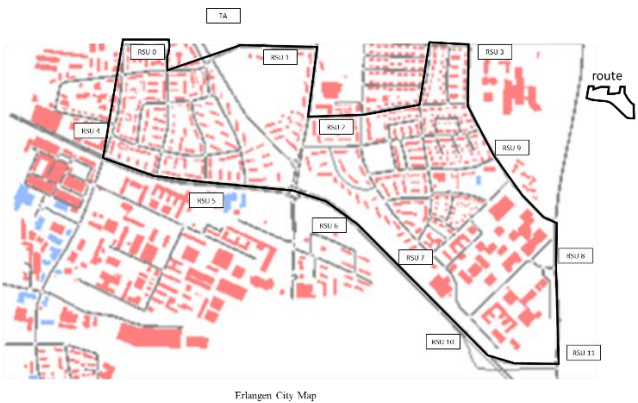

Fig. 13. Erlangen city map from [40].

### B. Behavioural Analysis of the Sender and Reporter Drivers

*1) Uniform Trust Distribution (0.4 to 0.5):* In this experiment, all vehicles are inserted, and drivers are assigned their initial trust using a uniform distribution in the range of 0.4 to 0.5. Fig. 14 records the lying behaviour data from this experiment. The x-axis shows the simulation seconds, and the y-axis shows how trust score changes from the rewards and punishments. Though 100 drivers are present, the trust records of most drivers are not included in this chart for simplicity as their trust remains constant.

*a) Results:* There is an accident message scheduled from V0 which is not announced as the trust of the driver is insufficient. This is why a change in the trust data only commences from 700s when V[0] announces a debris message. As the trust of V[0] is low, driver has a higher chance to lie to others which is modelled using a probabilistic distribution. As the driver of V[0] lies, the drivers of V[2] and V[3] improve their trust by sending untrue attacks and they win against the driver of V[0]. This is visible from the chart. The other two drivers do not participate in the reporting process and hence their trust remains constant over the simulation period. Also, V[5] wins one dispute over V[0] which is indicated by a trust increment at about 3600s. It is seen that the announcement of trustworthy messages varies based on a driver's trust state.

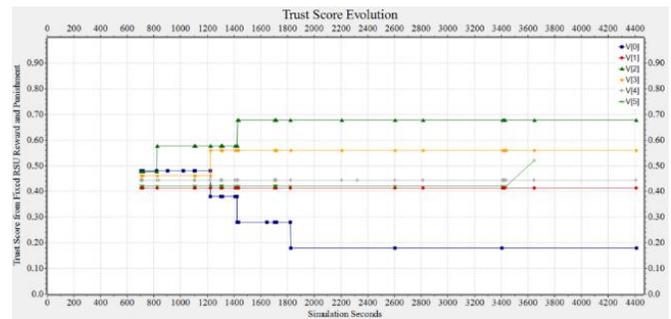

Fig. 14. Behavioural analysis of the drivers with trust (0.4-0.5).

*2) Fixed trust score of 0.9:* In this experiment, all drivers start from a very high trust state with a trust score of 0.9. Fig. 15 records the lying behaviour data from this experiment. The driver of V[0] is set to send 90% trustworthy and 10% of malicious announcements from this state.

*a) Results:* It is seen very few announcements are reported from V[1] and V[5] as they are also assigned "very good" trust states though their malicious probability is 0.2. This results in the constant trust score of the driver of V[0] while some reporters send untrue attacks maliciously which are disproved at RSUs. Hence, some reporters receive RSU punishments at different times during the latter part of the simulation. The drivers of V[1] and V[2] send only untrue attacks for which their trust is reduced. Thus, as configured, with a higher trust state there are fewer untrusted messages announced. Additionally, reporter drivers send fewer untrue attacks when their trust scores and corresponding trust states are higher.

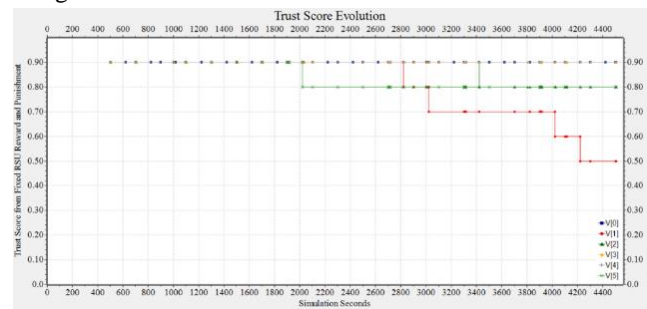

Fig. 15. Behavioural analysis of the drivers with trust=0.9.

*C. Behavioural Analysis of Sender with Fixed Trust (0.6) of Reporter and Clarifier*

*1) Simulation setup:* The set of parameters are same as they are listed in Table VII. 100 vehicles are added and then they elapse a warm-up period. One sender driver of V[0] sends message periodically and five reporters from V[0],…,V[5] send reports based on the probability distribution defined in Table VI. Table VIII lists the feedback generation probability of clarifier vehicles. In the next two experiments, clarifiers send feedback based on the probability distribution of their trust states. Also, the reporters send report based on the probability distribution of their trust states. After that their behaviour are captured in Fig. 16 and Fig. 17.

TABLE VIII. CLARIFIER'S FEEDBACK PROBABILITY

| Trust States | Probability of Sending Positive Feedback | Probability of Sending Negative Feedback |
|---|---|---|
| "very good" | 0.8 | 0.2 |
| "good" | 0.6 | 0.4 |
| "normal" | 0.4 | 0.6 |
| "bad" | 0.2 | 0.8 |
| "very bad" | 0.1 | 0.9 |
| "access blocked" | 0 | 0 |

*2) With sender driver's trust of 0.3:* Fig. 16 shows the trust score evolution of six vehicles. In this experiment, the trust of sender driver is set to 0.3, the trust of the reporter and the clarifier is set to 0.6. Clarifiers send opinion when an RSU asks based on their probability distribution of trust states. As, the trust score of the reporter and clarifier belong to the "Good" trust state. With these settings, reporter vehicles send untrue attacks with only 30% of cases and clarifiers send positive opinion in 60% of cases when they observe event on road. They also send negative feedback with 0.4 probability if they do not see the event on the said location. In this way, their communication is achieved.

*a) Results:* Until first 1400s, there is no dispute, and no trust change observes. After that, there are many reports announced for which the driver of V[0] only wins. As the reporters sends report maliciously. The reporter driver of V[5] loses all disputes which reduces trust to 0.2 at 2400s. The driver of V[3] does not report any announcements from V[0] until 2800s as seen from the chart. After this time, V[3] sends many reports which are proved false to RSU, so the trust is reduced to 0.2 at 4030s. Other reporters excluding V[4] occasionally sends untrue attacks and they also lose the disputes to V[0]. In this way, V[0] builds trust as always it announces trustworthy messages and some reporters being malicious lose trust. In Fig. 16, the sender driver slowly improves trust from only the RSU rewards, and the malicious reporters receive only RSU punishments as their reports are proved false by RSUs. As the sender is trustworthy throughout the simulation, all reporters receive RSU punishments which reduces their trust score, and their trust state moves from "normal" to "bad" and then "very bad" as a consequence.

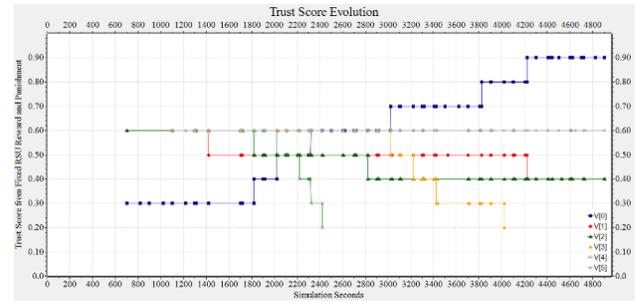

Fig. 16. Behaviour analysis of driver when trust score is 0.3.

*3) With sender driver's trust of 0.7:* In this experiment, sender driver starts with 0.7 trust score and from "normal" trust state whereas the clarifier's and reporter's trust state are same to the previous experiment. They both start with the "Good" trust state.

*a) Results:* In this experiment, the driver of V[0] only builds trust as always send trustworthy announcements. Reporter drivers V[2], V[5], and V[3] send reports maliciously for which they lose all disputes. These are noticed by the trust decrements in Fig. 17. Reporter V[4] does not send any report and V[1] sends only one untrue report for which it receives the RSU punishment. When a reporter sends a malicious report and receives RSU punishment, then subsequently it sends more report maliciously as their trust states moves toward "bad" state. As the sender driver reaches "Good" trust state early in Fig. 17, it only announces trustworthy messages and when reporters send false reports, they receive RSU punishments. As they move from "Normal" to "Bad" states the reporters send more false reports and hence they receive more RSU punishments.

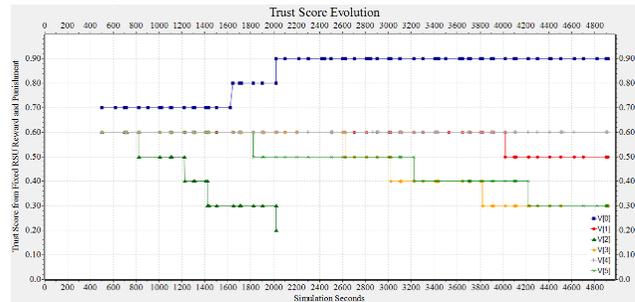

Fig. 17. Behaviour analysis of driver when trust score is 0.7.

VI. PERFORMANCE COMPARISON OF THE PROPOSED FUZZY JUDGEMENT VS. FIXED RSU JUDGEMENT

*A. Performance Comparison Using Only RSU Reward and Punishment*

*1) Simulation setup:* The fuzzy reward or punishment method is applied when dispute decisions are ready at RSUs. Hence, the comparison is made between the fuzzy vs fixed reward and punishment schemes. To this end, a series of experiments is conducted to evaluate their performance. When the warm-up period has elapsed, a fixed sender driver of V[0] announces messages periodically at 1000s periodic intervals for each event type starting from the 500s. In this set of experiments, multiple types of event announcement are considered from the same driver V[0] for this comparison which is announced similarly in the behaviour analysis. When they are announced, a fixed set of reporter drivers of vehicles V[1], V[2], V[3], V[4], and V[5] deterministically sends untrue attack reports after their reception. The trust data is recorded separately for both the fuzzy and fixed systems. The trust framework has other TPD rewards and punishments which are omitted for differentiating these assessment results since fuzzy logic is only used to improve the RSU reward and punishment mechanism. Updates to trust from the fuzzy and fixed reward and punishment schemes are shown on graphs to compare them. After this, two trust density distributions are presented for each scheme. One shows the initial trust data, and the other provides the trust data when the simulation ends.

*2) Scenario 1 – Trust Updates from the Fuzzy RSU:* Fig. 18 shows the trust score evolution for six vehicles only. Trust is updated only from RSU judgements. The x-axis represents simulation seconds, and the y-axis shows the updated trust from the fuzzy RSU unit. During this experiment, the driver of V[0] sends scheduled events periodically. The initial trusts are assigned from a uniform distribution with the range of 0.5 to 0.6. The driver of V[0] starts with a "normal" trust state which governs his/her behaviour in message announcements. This state is configured to send more malicious messages than the trustworthy messages in the state transition model.

*a) Results:* It is seen that V[0] builds trust from the fuzzy rewards as it announces only trustworthy messages while the reporters get fuzzy punishments which reduces their trust as the simulation progresses. First, V[0] moves to "Good" state and then to "Very Good" states. V[0] reaches the maximum trust at about 1800s with "Very Good" state. Alternatively, reporters in this experiment send untrue reports and move from the "normal" to the "bad" trust state. For example, the driver of V[2] always sends false reports and receives RSU punishments. His/her trust score plunges to the lowest value of 0.34 at 2900s due to being malicious. It is noticeable that the first reward of V[0] is highest as the driver has no punishment records in the DBP whereas the latter judgements are not seen as high as the first one. Since, some latter rewards are from the disputes relating to the less severe announcements. Alternatively, the fuzzy RSU punishments are not very harsh initially which is seen in the reporter vehicles V[1] and V[5]'s punishments. They increase slightly in the later punishments where the severity of incident, punishment records in the DPB, and RSU confidence influence the outcome. In later disputes, event severity levels are different which vary the punishment. Hence, the rewards / punishments vary throughout the experiment whereas in the fixed reward scheme trust increments / decrements are fixed irrespective of mitigating factors. So, with the fuzzy scheme, a driver has more chances to improve trust scores from subsequent announcements and trustworthy reporting. This way their network participation lifespan is extended. Fig. 19 depicts the trust scores of all vehicles which participated in this experiment. It is noticeable from this figure that the trust scores of most vehicles are unchanged throughout the simulation as they do not report or announce any messages and there is no forwarding or clarifying reward for others.

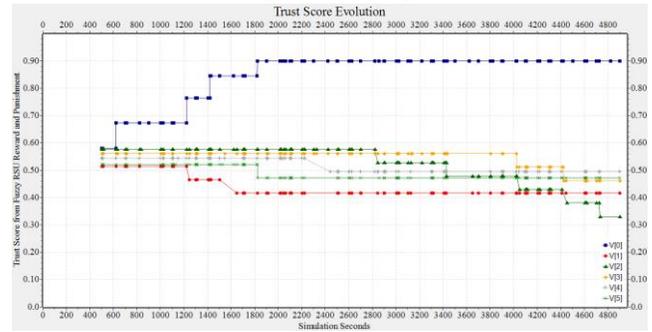

Fig. 18. Trust score evolution of six vehicles from the fuzzy reward and punishment.

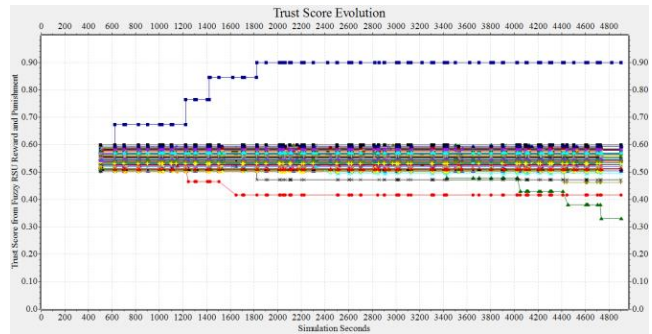

Fig. 19. Trust score evolution of all vehicles with the fuzzy reward and punishment.

Fig. 20 shows the trust score density distribution of vehicles collected at the beginning and at the end of the simulation. The initial trust score of all vehicles is between 0.5 and 0.6. The right-side chart shows that V[0] reaches 0.9 which is marked by a dot. Most vehicles do not see any trust score alterations apart from three vehicles which are the reporters in this experiment. This is because general vehicles do not engage in any disputes from which they can earn or lose trust. Additionally, they are not given any reward from the forwarding or other activities. The long gap in the right chart means no vehicle other than V[0] achieves this score due to the experimental design and this result is as expected. Also, the driver behaviour model governs their honest and dishonest announcements. It is seen with the fuzzy system, the magnitude of reward and punishment are more nuanced than the fixed system so vehicles have more time to correct their future behaviour and resume normal operation. The blacklisting of a vehicle or reaching the highest trust is also delayed when using the fuzzy system. Even so, in the fuzzy system when only RSU rewards and punishments are given, the sender vehicle still reaches 0.9 trust. The reporter vehicle reaches a low trust score though it has some trust left to carry out further communication and it could choose to correct its behaviour and achieve good trust score in due course. Overall, with the fuzzy system the trust scores are more stable than the fixed system in the sense that when trust is gained or lost it does not change dramatically. Additionally, the fuzzy system considers environmental dynamics for fuzzy judgements, e.g., event severity, driver past behaviour and confidence score which is appropriate when reviewing disputes.

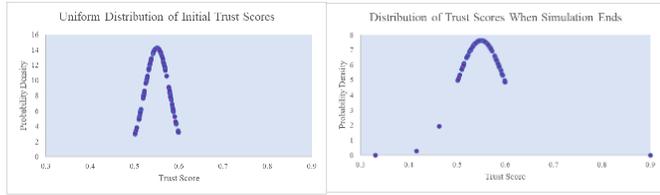

Fig. 20. Distribution of trust scores at the start and the end.

*3) Scenario 2 – Trust Updates from the Fixed RSU:* The next experiment measures the trust score of vehicles only from the fixed RSU reward and punishment mechanism. In Fig. 21, simulation time is on the x-axis and the y-axis shows the trust score. This is conducted with the set of parameters defined in Table V, but the RSU reward and punishment is fixed (0.1) for every driver.

*a) Results:* V[0] sends a malicious message initially and receives an RSU punishment that reduces its trust below 0.5. From this stage, it is configured to send more malicious messages 80% of time. Thus, its trust subsequently decreases from RSU punishments. When its trust score belongs to the "very bad" state, it sends all malicious messages. In this way, its trust is reduced to 0.05 which meets the condition to block its access. Alternatively, the reports from V[1] wins all disputes and hence its characteristic shows an upward trend. Also, V[4] and V[5] win two other disputes over V[0] and hence receive RSU rewards. It should be noted that there are no events after the 4400 seconds. It is seen that trust adjustments are faster in the fixed RSU judgement system as it assigns a higher amount (0.1) irrespective of event type and driver behaviour compared to the fuzzy system which provides a value in the range 0.01 to 0.1 based on the evaluation result. When only RSU rewards and punishments are given, in the fixed system vehicle V[0] is access-blocked. This is due to the RSU decisions about the announcement being untrue along with the magnitude of the penalty. Fig. 22 shows the trust scores of all vehicles in this experiment. In this figure most of the vehicles do not change their trust score as they do not participate in any reporting or announcement. Besides, they are not given any reward for clarifying and forwarding.

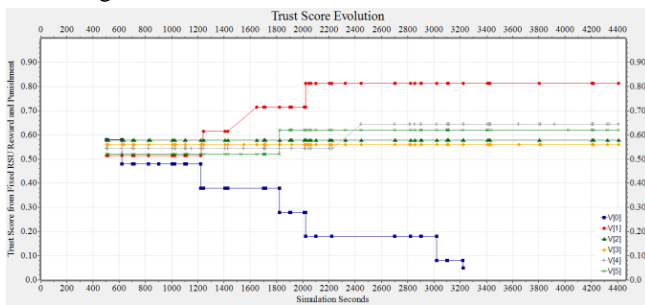

Fig. 21. Trust score evolution from fixed reward and punishment.

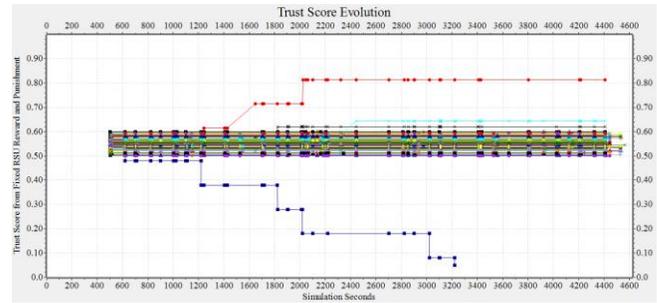

Fig. 22. Trust score evolution with fixed reward and punishment.

Fig. 23 shows the initial trust and the final trust distribution in two density curves. The first density chart shows the trust scores of all vehicles generated from a uniform distribution. However, the right-hand chart plots the trust scores of all vehicles when the simulation ends. As expected, in the second chart, the trust of most vehicles is unchanged as they do not engage in any disputes from which their trust can change. The right-hand chart confirms some vehicles with positive behaviour build their trust from truthfully reporting activities whereas the sender V0 is access-blocked, leaving its trust at 0.05. With this fixed RSU judgement, vehicles have less opportunity to modify their behaviour and vehicle access-blocking is more likely as shown in the right chart in Fig. 21.

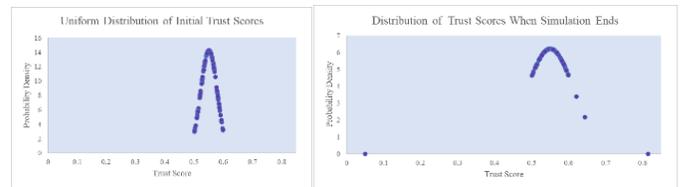

Fig. 23. Distribution of trust scores at the beginning and the end.

## VII. DISCUSSION

The trust model with the fuzzy reward or punishment assessment scheme controls the broadcasting of messages at the sender side based on the trust score of drivers / vehicles so a receiver driver / vehicle can believe in a message instantly and does not need to take any further action. By regulating the ability to broadcast, malicious vehicles, once identified, are unable to continue to broadcast messages. Though, blacklisting is present in most existing approaches it requires the trust score to reach zero. Therefore, a malicious vehicle can create many hazardous problems before being blacklisted. Furthermore, driver decision times (response times) are reduced as trust does not need to be verified on a per message basis. This also reduces the communication overhead. It uses the RSU for ruling on a dispute when needed using clarifier feedback, rather than approaches which gather direct and/or indirect trust or opinions from surrounding neighbours, which may include false recommendations from malicious vehicles. However, the application of fuzzy reward or punishment assessment is only limited to the dispute resolution process at the RSU. Additionally, it requires presence of clarifier vehicles to send feedback to assist in the resolution of disputes.

## VIII. CONCLUSION AND FUTURE WORK

In this paper, a Mamdani fuzzy logic based RSU reward and punishment assessment scheme is presented. This application considers event severity, driver past behaviour and RSU confidence (calculated from the feedback of the clarifier vehicles) to determine an appropriate level of reward or punishment for the drivers involved. The reward and punishment mechanism uses a different set of rules to assess the output. The fuzzy RSU reward or punishment assessment scheme is an extension of the fixed RSU judgement mechanism in the previous sender-side trust framework [6]. The RSU ruling is only needed when there is a dispute (when both an event and opposite event exist) in the network. The fuzzy RSU controller is invoked only when it receives untrue attack reports, which is expected to be occasional. A Markov-chain based driver behaviour model is also included to control the announcement behaviour of driver when conducting the series of experiments.

The fuzzy approach is compared against a fixed reward and punishment scheme. Trust evolution timelines are provided in each case along with trust density distribution curves when only the RSU mechanism is active. The results suggest the fuzzy system achieves a more stable trust environment. This assessment also employs a Markov-chain based driver behaviour model whereby good drivers are assumed to behave in a positive manner more generally, and vice versa. This allows the nuanced fuzzy controller decisions to encourage drivers to behave better, and to provide fairer allocation of rewards and punishments based on several factors. However, in the future other inputs to the fuzzy controller could be considered.